# Combining local search techniques and path following for bimatrix games

**Nicola Gatti, Giorgio Patrini, Marco Rocco**
Dipartimento di Elettronica e Informazione
Politecnico di Milano
Milano, Italy

**Tuomas Sandholm**
Computer Science Department
Carnegie Mellon University
Pittsburgh, USA

## Abstract

Computing a Nash equilibrium (NE) is a central task in computer science. An NE is a particularly appropriate solution concept for two–agent settings because coalitional deviations are not an issue. However, even in this case, finding an NE is PPAD–complete. In this paper, we combine path following algorithms with local search techniques to design new algorithms for finding exact and approximate NEs. We show that our algorithms largely outperform the state of the art and that almost all the known benchmark game classes are easily solvable or approximable (except for the GAMUT CovariantGame–Rand class).

## 1 Introduction

Finding a Nash equilibrium (NE) of a normal–form (aka strategic–form aka bimatrix) game is PPAD–complete [8] even with just two players [6]. Although PPAD $\subseteq$ NP (but PPAD $\not\subseteq$ NP–complete unless NP = co–NP) it is generally believed that PPAD $\neq$ P. Thus the worst–case complexity of finding an NE is exponential in the size of the game. This pushed researchers to study approximate solution concepts, providing some polynomial–time algorithms that compute solutions with a guaranteed approximation bound [2, 11, 22]. However, bimatrix games do not have a fully polynomial–time approximation scheme unless PPAD $\subseteq$ P [6].

*Worst–case* complexity, being too pessimistic, is often a bad indicator of the actual performance of an algorithm, and *average–case* complexity is difficult to determine. A new metric of complexity, called *smoothed* complexity, has been gaining interest in recent years [21]. It studies how the introduction of small perturbations affects the worst–case complexity. Unfortunately, finding an NE in two–player games is not smoothed polynomial unless PPAD $\subseteq$ RP [5].

In this paper, we focus on two–player general–sum strategic–form games. An NE is a particularly appropriate solution concept in two–agent settings because coalitional deviations are not an issue. The main known algorithms are LH [14] based on linear complementarity mathematical programming (LCP), PNS [18] based on support enumeration, and MIP Nash [19] based on mixed–integer linear programming (MILP). None of these beats the others on all games, and all of them have worst–case complexity exponential in the size of the game. In particular, there are game instances, called hard–to–solve games (HtSG), that always require exponential time when solved with LH and PNS [20].

The integration of local search techniques with support enumeration algorithms, called LS–PNS, can be effective on games that are hard for all three algorithms above [3, 4]. (Other local search algorithms, based on simulated annealing or homotopy, that are slower but have convergence guarantees, have also been designed [12, 13].) In this paper, we combine local search techniques with path following algorithms (e.g., LH) instead. This opens new promising opportunities, allowing a dramatic reduction of the solution space: path following algorithms work on a solution space that is $O(2.6^m)$ where $m$ is the number of actions per agent, while the number of supports is $O(4^m)$. Our contributions include the following.

($i$) We design new algorithms: ($a$) an LH version with random restarts (rrLH)[1], ($b$) a version of the adaptation of the Lemke algorithm (L) proposed in [24] with random restarts (rrL), ($c$) a local search algorithm moving on the best response vertices (LS–v), and ($d$) an anytime LH version based on iteratively decreasing perturbation of the payoff matrices (ip–LH).

---

[1] A similar algorithm is presented in [7]. We extend such algorithm with an iterative deepening cutoff, we theoretically analyze its properties, and we provide a more thorough experimental analysis, developing a floating–point implementation that is about two times faster and developing a new arbitrary–precision implementation.

(ii) We prove that rrLH is asymptotically optimal in the space of algorithms that randomize over LH paths (the result can be extended to rrL).

(iii) We experimentally show that LH requires arbitrary precision with some classes of GAMUT games [17] and HtSG. L requires arbitrary precision for almost all the game instances.

(iv) rrLH outperforms all our and state–of–the–art algorithms, requiring an average number of steps that is linear in the size of the game except for CovariantGame–Rand and HtSG. rrL works asymptotically as rrLH, but its compute time is usually larger.

(v) LS–v outperforms the other algorithms in approximating hard instances, except HtSG.

(vi) HtSG can be easily approximated by ip–LH even with a perturbation of $10^{-10}$. Thus there must exist a different worst case for LH unless PPAD $\subseteq$ RP. (In contrast, CovariantGame–Rand stays hard even when perturbed, thus it is a possible candidate.)

## 2 Game model and solution concepts

A bimatrix game is a tuple $(N, A, U)$, where $N = \{1, 2\}$ is the set of agents (we denote by $i \in N$ a generic agent); $A = (A_1, A_2)$ where $A_i$ is the set of $m_i = |A_i|$ actions available to agent $i$ (we denote by $a \in A_1 \cup A_2$ a generic action); $U = (U_1, U_2)$ where $U_i$ is the utility matrix of agent $i$ [9]. Without loss of generality, we assume that $\max_{j,k}\{U_i(j,k)\} = 1$ and $\min_{j,k}\{U_i(j,k)\} = 0$ for every $i \in N$.

We denote by $\mathbf{x}_i$ the strategy (vector of probabilities) of agent $i$ and by $x_{i,a}$ the probability with which agent $i$ plays action $a \in A_i$. We denote by $\Delta_i$ the space of strategies over action space $A_i$, i.e., vectors where the probabilities sum to 1. Given strategy $\mathbf{x}_i$, the *support* $S_i$ is the set of actions $a \in A_i$ played with a non–zero probability in $\mathbf{x}_i$.

The central solution concept is NE. A profile of strategies $(\mathbf{x}_1, \mathbf{x}_2)$ is an NE if and only if, for each $i \in N$, $\mathbf{x}_i^T U_i \mathbf{x}_{-i} \geq \mathbf{x}_i'^T U_i \mathbf{x}_{-i}$ for every $\mathbf{x}_i' \in \Delta_i$. The problem of finding an NE can be expressed as an LCP:

$$\mathbf{x}_i \geq 0 \quad \forall i \in \{1,2\} \quad (1)$$
$$\mathbf{1}v_i - U_i \mathbf{x}_{-i} \geq 0 \quad \forall i \in \{1,2\} \quad (2)$$
$$\mathbf{x}_i^T (\mathbf{1}v_i - U_i \mathbf{x}_{-i}) = 0 \quad \forall i \in \{1,2\} \quad (3)$$
$$\mathbf{1}^T \mathbf{x}_i = 1 \quad \forall i \in \{1,2\} \quad (4)$$

Here $v_i$ is the expected utility of agent $i$. Constraints (1) and (4) state that every $\mathbf{x}_i \in \Delta_i$. Constraints (2) state that no pure strategy of agent $i$ gives expected utility greater than $v_i$. Constraints (3) state that each agent plays only optimal actions.

The most common approximate solution concept is $\epsilon$–Nash equilibrium ($\epsilon$–NE): $(\mathbf{x}_1, \mathbf{x}_2)$ is an $\epsilon$–NE if, for each $i \in N$, $\mathbf{x}_i^T U_i \mathbf{x}_{-i} \geq \mathbf{x}_i'^T U_i \mathbf{x}_{-i} - \epsilon$ for every $\mathbf{x}_i' \in \Delta_i$. Informally, $(\mathbf{x}_1, \mathbf{x}_2)$ is an $\epsilon$–NE if at most each player loses a utility of $\epsilon$ w.r.t. playing her best response. Other approximate solution concepts that we use in this paper are $\epsilon$–Well Supported–Nash equilibrium ($\epsilon_{WS}$–NE) and regret–Nash equilibrium ($r$–NE). $(\mathbf{x}_1, \mathbf{x}_2)$ is an $\epsilon_{WS}$–NE if, for each $i \in N$, $\mathbf{e}_k^T U_i \mathbf{x}_{-i} \geq \mathbf{e}_j^T U_i \mathbf{x}_{-i} - \epsilon$ for every $k \in S_i, j \in A_i$ ($\mathbf{e}_k$ is a vector of 0 but 1 in position $k$). Thus playing singularly each action of the support, the agent loses a utility of at most $\epsilon$ w.r.t. playing the her best response. Finally, $(\mathbf{x}_1, \mathbf{x}_2)$ is an $r$–NE if, for each $i \in N$, $r = \sum_{i=\{1,2\}} \sum_{l \in S_i} r_{i,l}$ and $\mathbf{e}_k^T U_i \mathbf{x}_{-i} \geq \mathbf{e}_j^T U_i \mathbf{x}_{-i} - r_{i,k}$ for every $k \in S_i, j \in A_i$.

## 3 Randomizing over paths

In this section we develop NE algorithms that randomize over *almost complementary* paths.

### 3.1 Randomizing over Lemke–Howson (LH) paths

For every agent $i$, define the best response polyhedron $P_i = \{\tilde{\mathbf{x}}_i \in \mathbb{R}^{m_i} | U_{-i}\tilde{\mathbf{x}}_i \leq \mathbf{1}, \tilde{\mathbf{x}}_i \geq \mathbf{0}\}$. Let $V_i$ be the vertices of $P_i$. The space $\Theta$ of solutions traversed by LH is a subset of pairs of vertices of $V = V_1 \times V_2$. Let $\mathbf{s}_{-i}$ be the slack variables of $U_{-i}\tilde{\mathbf{x}}_i + \mathbf{s}_{-i} = \mathbf{1}$. Variables $s_{i,a}$ and $x_{i,a}$ are called *complementary*. A basic solution of the tableaux associated with $P_1$ and $P_2$ is *complementary* if the basis contains exactly one complementary variable for all $i$ and $a$, while it is *almost complementary* if both variables of a single pair of complementary variables are in the basis. A completely complementary solution is an NE.

The initial solution of the LH algorithm [14] is generated by starting from the artificial solution $(\mathbf{0}, \mathbf{0})$ (corresponding to the case in which all the slack variables $\mathbf{s}_i$, for every $i$, are in the basis), and putting in the basis of one of the two tableaux a variable $x_{i,a}$ associated with some action $a \in A_1 \cup A_2$. Thus, there are $m_1 + m_2$ different possible initial solutions. LH follows a path of almost complementary solutions by repeatedly applying complementary pivoting steps. At each step, the current basis is changed by putting in the basis the variable that is complementary to the variable that has left the basis in the previous step, until a completely complementary solution has been found. The leaving variable is determined by the minimum ratio test [15] and it is unique except for degenerate games; however, degeneracy can be removed by introducing lexicographic perturbation [24].

For each initial solution, there is a different path leading to a (potentially different) NE: LH partitions the solution space in $m_1 + m_2$ different paths. If the path that the algorithm is following is an exponentially long one, it may be convenient to have restart.

We observe that any random restart policy over the solutions traversed by the LH algorithm can be formulated as a two–stage randomization policy: (1) ran-domization over the $m_1 + m_2$ possible paths and (2) randomization, given a path, over its solutions. We initially present an algorithm, called rrLH, that adopts only a blind randomization of type (1) and, subsequently, we investigate randomization of type (2).

rrLH is described in Algorithm 1. It randomly chooses one of the paths (Step 2) and follows it (Step 4) until an NE is found; if the length (in terms of number of pivoting steps) of the path is larger than a given *cutoff* and there is a path that has not been visited till *cutoff* (Steps 5, 6, and 8), the algorithm restarts with a new path, otherwise (Step 7) *cutoff* is updated in an iterative deepening fashion and a restart is done.

**Algorithm 1** Lemke–Howson with random restart (rrLH)

1: $cutoff = cutoff_0$
2: randomly choose one path non–visited till *cutoff*
3: **repeat**
4:   apply complementary pivoting
5:   **if** the path is longer than *cutoff* **then**
6:     **if** all the paths are visited till *cutoff* **then**
7:       $cutoff = cutoff + cutoff_0$
8:     go to Step 2
9: **until** a completely complementary solution is found
10: return current solution

(To improve the efficiency of the algorithm, we save the variables of the current basis before making a restart and, when a path is re–visited, we derive the last visited basis by matrix inversion and we restart from it.) The advantages of rrLH are simplicity and completeness. The potential drawbacks are that the number of possible paths is limited and that the algorithm is forced to move along fixed paths. Call $l^*$ the shortest LH path. The compute time of rrLH is linear in $l^*$. When $l^*$ is known or it is possible to estimate a tight upper bound, $cutoff_0$ can be conveniently fixed. Assigning, e.g., $cutoff_0 = l^*$, we have that the worst case compute time is $l^*(m_1 + m_2)$, while the average is $l^*(m_1 + m_2 + 1)/2$. Thus, if $l^*$ grows in length exponentially in $m_1 + m_2$, also rrLH compute time grows in length exponentially. Adopting a non–blind randomization of type (1) would keep the compute time, even in the best case, linear in $l^*$ and hence is useless.

We focus on randomization of type (2). This type of randomization cannot be efficiently achieved with LH. Indeed, differently from what happens with support–enumeration algorithms [3, 4] where it is possible to start from any support profile, no every possible almost complementary basis corresponds to a feasible starting solution for LH: some sets of almost complementary variables are not feasible basic solutions, while the feasible ones can belong to LH paths or not, and in this latter case they may belong to cyclic paths that do not lead to any equilibrium (the membership of a solution to a path can be discovered only during the traversing of the path itself and therefore, if the algorithm starts from an arbitrary solution, it cannot know the path over which it is moving and cannot distinguish paths from cycles). Thus, an initial solution must be searched by using pivoting and the number of pivoting steps may be exponential in $m_1 + m_2$. However, we show that any algorithm with randomization of type (2) has a compute time that is asymptotically the same of our rrLH and therefore rrLH is (asymptotically) optimal in the space of the algorithms making random restarts over the LH paths. This holds even dropping completeness and considering algorithms that find an NE with a probability $p$. Initially, we state the following lemma.

**Lemma 1.** *The best* cutoff *of a generic randomized algorithm that finds the terminal vertex of an $l$–step–long path with probability $p$ and is able to position blindly in every vertex of the path is $l \cdot p$.*

*Proof.* The best configuration of the algorithm can be obtained by minimizing the number of steps of the algorithm (i.e., *cutoff* · *res*, where *res* is the number of restarts), under the constraint that the probability to find the terminal is $p$, i.e., $p = 1 - (1 - cutoff/l)^{res}$, $res \geq 1$, and $cutoff \geq 1$. From $p = 1 - (1 - cutoff/l)^{res}$ it follows that $res = \log(1-p)/\log(1-cutoff/l) \geq 1$, thus $\log(1-p) \geq \log(1-cutoff/l)$, that means $p \leq cutoff/l$. $res \cdot cutoff$ can be rewritten as $\log(1-p)/\log(1-cutoff/l) \cdot cutoff$. After having removed the negative constant $\log(1-p)$ the objective is to maximize the monotonic decreasing function $cutoff/\log(1-cutoff/l)$ under the constraint $p \leq cutoff/l$. The optimum is $cutoff = l \cdot p$ and $res = 1$. □

From Lemma 1 it follows that, even when it is possible to perform a blind randomization over the solutions composing a single path, like stage (2) prescribes, the optimal configuration of the algorithm is such that this path is traversed only once without making restarts (i.e. $res = 1$). Thus, this path can be safely removed from the set of the available paths at stage (1). From Lemma 1, we can easily derive the following lemma.

**Lemma 2.** *The worst case compute time of a randomized algorithm finding the terminal vertex of an $l$–step–long path and able to position blindly in every vertex of the path is $l$, while the average time is $l/2$.*

Finally, we show that including randomization of type (2) the compute time keeps to be linear in $l^*$ as stated by Lemma 3. From Lemma 2 we can state the following.

**Lemma 3.** *LH with blind randomization policy of type (2) has a compute time $O(l^*)$.*

We can evaluate for specific interesting cases the ratio between the average compute time of an algorithm $\Pi$ adopting randomization of type (2) and the one of an algorithm $\Pi'$ that does not. Assign $cutoff = l^*$ and assume, for simplicity, that the length of all the non–shortest paths is $l > l^*$ and $m_1 = m_2 = m$.

The average compute time of $\Pi$ is $(2m + 1)/2 \cdot l^*$. The average case compute time of $\Pi'$ is given by $1/2m \cdot l^*/2 \cdot \sum_{k=0}^{2m}(1-l^*/2)^k + (l^*)^2/2l \cdot 1/2n \cdot \sum_{k=0}^{2m-1}(1-l^*/l)^k(2m-1-k)$. The two opposite possible situations are: $l^*/l \to 1$ and $l^*/l \to 0$. In the best situation $l^*/l \to 1$ the average compute time is $l^*/2$ and therefore randomization of type (2) reduces the compute time by $2m+1$ times. In the worst situation $l^*/l \to 0$, the ratio is 1 and therefore no reduction is obtained. In both cases, the compute time keeps to be linear in $l^*$. We observe that, randomization of type (2) exploits the existence of paths that are not excessively longer than the shortest path.

Now, we focus on non–blind randomization of type (2), providing a negative result.

**Lemma 4**. *Any algorithm that finds the terminal vertex of an $l$–step–long path with probability $p$ and able to position non–blindly in every vertex of the path, requires either an exponential number (res) of restarts or an exponential* cutoff *as $l$ grows in length exponentially.*

*Proof.* Suppose to have an oracle that, given a *cutoff* and an almost complementary solution, is able to say whether or not such a solution is farther than *cutoff* from the terminal vertex. If the solution is farther, then it is discarded, otherwise the algorithm follows the path from the given solution and the terminal vertex. The probability to find a randomly generated solution that is not farther than *cutoff* from the equilibrium by *res* random restarts is: $1-(1-cutoff/l)^{res}$. By posing: $1-(1-cutoff/l)^{res} = p$, we obtain $res = \log(1-p)/\log(1-cutoff/l)$. When $\lim_{l\to+\infty} cutoff/l = 0$, we can write $res = -l\log(1-p)/cutoff$. Therefore, if $l$ grows in length exponentially then either *res* or *cutoff* grow in length exponentially. When $\lim_{l\to+\infty} cutoff/l > 0$, *cutoff* grows in length exponentially as $l$ does. □

From the above lemma, it can be easily observed that when all the LH paths grow in length exponentially non–blind randomization is useless: even dropping the completeness and accepting that the NE can be found with a probability $p$, in the worst case the compute time is $O(l^*)$. Thus, rrLH is asymptotically optimal among algorithms randomizing over LH paths.

### 3.2 Randomizing over Lemke (L) paths

As discussed in the previous section, randomization over paths exhibits a compute time that depends on the length of shortest path. The main drawback of LH is that the number of available paths is small. In this section, we resort to the Lemke algorithm adaptation as prescribed by [24], which we will call L. This algorithm allows for an arbitrary initial solution, each corresponding to a different path, and therefore it allows for an infinite number of paths. Define the polyhedron $P$ as follows.

$$P = \left\{ [z_0 \ \mathbf{z}_1 \ \mathbf{z}_2] \middle| \begin{array}{l} M_{1,1}\mathbf{z}_1 + M_{1,2}\mathbf{z}_2 + \mathbf{d}_1 z_0 = \mathbf{q}_1 \\ M_{2,1}\mathbf{z}_1 + M_{2,2}\mathbf{z}_2 + \mathbf{d}_2 z_0 + \mathbf{q}_2 \geq 0 \\ z_0, \mathbf{z}_2 \geq 0 \end{array} \right\}$$

with $\mathbf{d}_1 = \mathbf{1}, \mathbf{d}_2 = [-(U_1\overline{\mathbf{x}}_2)^T, -(U_2\overline{\mathbf{x}}_1)^T]^T$ where $\overline{\mathbf{x}}_1$ and $\overline{\mathbf{x}}_2$ are parameters (therefore $P$ is parametric), $\mathbf{q}_1 = -\mathbf{1}$, $\mathbf{q}_2 = \mathbf{0}$ and

$$\mathbf{z}_1 = \left[\begin{array}{c} v_1 \\ v_2 \end{array}\right] \quad M_{1,1} = \left[\begin{array}{cc} 0 & 0 \\ 0 & 0 \end{array}\right] \quad M_{1,2} = \left[\begin{array}{cc} \mathbf{1}^T & \mathbf{0}^T \\ \mathbf{0}^T & \mathbf{1}^T \end{array}\right]$$

$$\mathbf{z}_2 = \left[\begin{array}{c} \mathbf{x}_1 \\ \mathbf{x}_2 \end{array}\right] \quad M_{2,1} = \left[\begin{array}{cc} \mathbf{1} & \mathbf{0} \\ \mathbf{0} & \mathbf{1} \end{array}\right] \quad M_{2,2} = \left[\begin{array}{cc} 0 & -U_1 \\ -U_2 & 0 \end{array}\right]$$

Let $V$ be the set of vertices of $P$. The space $\Theta$ of solutions traversed by the Lemke algorithm is a subset of $V$. Call $\mathbf{w}$ the slack variables of $M_{2,1}\mathbf{z}_1 + M_{2,2}\mathbf{z}_2 + \mathbf{d}_2 z_0 + \mathbf{q}_2 - \mathbf{w} = 0$ and consider the associated tableau. Call $z_j$, with $j \neq 0$, the $j$–th element of $\mathbf{z} = [\mathbf{z}_1^T, \mathbf{z}_2^T]$. Variables $z_j$ and $w_j$ are called complementary. A solution is completely complementary if the basis contains one complementary variable between $z_j$ and $w_j$ for every $j$ (therefore $z_0$ is out the basis), while a solution is almost complementary if both variables $z_j$ and $w_j$ for a single $j$ are not in the basis but $z_0$ is.

The algorithm moves along almost complementary solutions, each corresponding to a pair of strategies $(\mathbf{x}_1 + z_0\overline{\mathbf{x}}_1, \mathbf{x}_2 + z_0\overline{\mathbf{x}}_2)$. The initial solution is such that: $(a)$ $z_0$ is in the basis and it is equal to 1, $(b)$ all the variables $x_{i,a}$ such that $a$ is a best response to $\overline{\mathbf{x}}_{-i}$ are in the basis except one, and $(c)$ for all the non–best–response actions $a$ the complementary variable $w_j$ of $x_{i,a}$ is in the basis. Given the initial solution, the algorithm follows a path of almost complementary solutions by repeatedly applying complementary pivoting (the entering variable is the complementary variable of the leaving variable at the previous step, while the leaving variable is determined by the lexico minimum ratio test). At the first step of L, the entering variable is $x_{i,a}$ such that $a$ is the only best response not yet in the basis. (L terminates when $z_0$ is the leaving variable, finding as NE.)

LH has a finite number of possible paths $(m_1 + m_2)$, while L has an infinite number of them, thus we try to design a non–blind randomization policy among paths and we fixed a limit to the restarts, otherwise in the worst case the compute time would be infinite. We design the version of L with random restarts (rrL) like rrLH except that: $(a)$ the initial solution is determined by randomly generating $\overline{\mathbf{x}}_1$ and $\overline{\mathbf{x}}_2$; since there are infinitely many possible initial solutions and many of them could lead to potentially long paths, $(b)$ we use quality metrics to accept or discard an initial solution $\theta$ (we accept $\theta$ if $g(\theta) > th$, where $g$ can be defined in different ways, e.g., $\epsilon$, $\epsilon_{WS}$, and $r$, and $th$ is a threshold), $(c)$ we use a cutoff as a function of how the different metrics $\epsilon$, $\epsilon_{WS}$, $r$, and $z_0$ decrease along the path; and $(d)$ we disable the cutoff after a given number of restarts to guarantee completeness, the potential restarts would be infinite otherwise.

The advantage of rrL is that the initial solution can be any and thus much more paths than rrLH can be followed. The potential drawbacks are that the algorithm is forced to move along fixed paths and that the pivoting step requires about twice as much compute time as LH due to the tableau size.

## 4 Local search on best response vertices

We cast NE finding as a local search based optimization problem $(\Theta, f, N)$ where $\Theta$ is the solution space, $f$ is a function $f : \Theta \to \mathbb{R}$ to minimize, and $N$ is the *neighborhood function* that specifies, for each solution $\theta \in \Theta$, a set $N(\theta) \subseteq \Theta$ of solutions that can be directly reached from $\theta$ [16].

The solution space $\Theta$ is the set of vertices $V_i$ of the polyhedron $P_i = \{\tilde{\mathbf{x}}_i \in \mathbb{R}^{m_i} | U_{-i}\tilde{\mathbf{x}}_i \leq \mathbf{1}, \tilde{\mathbf{x}}_i \geq \mathbf{0}\}$ that constitutes the best response polyhedron of agent $i$, where agent $i$ is the agent with the minimum number of actions. This choice allows one to reduce the solution space as shown below (however, our algorithm can be applied with any $i$). Every vertex $\tilde{\mathbf{x}}_i$ of $V_i$, except for $\tilde{\mathbf{x}}_i = \mathbf{0}$, is equivalent to a strategy $\mathbf{x}_i = \frac{1}{\mathbf{1}^T \tilde{\mathbf{x}}_i} \tilde{\mathbf{x}}_i$.

The function $f$ to minimize associates each strategy $\mathbf{x}_i = \overline{\mathbf{x}}_i$ with the $\epsilon$ value of the best $\epsilon$–Nash equilibrium when agent $i$ plays $\overline{\mathbf{x}}_i$. This value can be computed as:

$$\min_{\epsilon, \mathbf{x}_{-i}} \quad \epsilon \tag{5}$$

$$\text{s.t.} \quad \overline{\mathbf{x}}_i^T U_i \mathbf{x}_{-i} + \epsilon - \mathbf{e}_k^T U_i \mathbf{x}_{-i} \geq 0 \quad \forall k \in A_i \tag{6}$$

$$\overline{\mathbf{x}}_i^T U_{-i} \mathbf{x}_{-i} + \epsilon - \overline{\mathbf{x}}_i^T U_{-i} \mathbf{e}_k \geq 0 \quad \forall k \in A_{-i} \tag{7}$$

$$\mathbf{1}^T \mathbf{x}_{-i} = 1 \tag{8}$$

$$\mathbf{x}_{-i} \geq \mathbf{0} \tag{9}$$

$$\epsilon \geq 0 \tag{10}$$

The constraints ensure that $\mathbf{x}_{-i} \in \Delta_{-i}$ and $\epsilon$ is non–negative. Note that the solution $\epsilon^*$ of the above linear program is 0 if and only if there is some $\mathbf{x}_{-i} = \overline{\mathbf{x}}_{-i}$ such that $(\overline{\mathbf{x}}_i, \overline{\mathbf{x}}_{-i})$ is an NE. Therefore, we recognize whether or not a local minimum is a global minimum.

Given a vertex $\tilde{\mathbf{x}}_i$ of $V_i$, the neighbors are all the adjacent vertices of $\tilde{\mathbf{x}}_i$. These vertices can be found by exploiting pivoting. More precisely, consider the tableau $U_{-i}\tilde{\mathbf{x}}_i + \mathbf{s} = \mathbf{1}$ where $\mathbf{s}$ are slack variables. A basic solution of the tableau is composed by $m_{-i}$ variables belonging to $\tilde{\mathbf{x}}_i$ and $\mathbf{s}$ (when only variables $\mathbf{s}$ are in the basis we have the artificial solution $\tilde{\mathbf{x}}_i = 0$). A vertex $\tilde{\mathbf{x}}_i$ of $V_i$ corresponds to a basic solution of the tableau. We can change the basis by making a non–basic variable enter the basis (the leaving variable is uniquely determined by the minimum ratio test). The number of entering variables is $m_i$; thus each vertex has exactly $m_i$ neighbors. In contrast, in the local search over supports, each solution has $m_1 \cdot m_2$ neighbors.

The local search algorithm for solving $(\Theta, f, N)$, called LS–v, is provided in Algorithm 2. The initial solution (Step 1) is generated starting from the artificial solution $\tilde{\mathbf{x}}_i = \mathbf{0}$ and applying a random number of random pivoting steps (as described above). In principle, any possible vertex of $P_i$ may be an initial solution. We use a tabu list to keep track of previously generated initial solutions in order to avoid repetitions. Given solution $s$, the algorithm searches in the neighborhood $N(s)$ for a solution $s'$ such that $f(s') < f(s)$ (Step 3). This search is driven by a heuristic. We use three main heuristics: *best improvement* (BI) where all neighbors are generated and the best neighbor better than the current is chosen, *first improvement* (FI) where the neighbors are searched in a given order and the first neighbor better than the current is chosen, *first improvement with random generation* (FIR) where the neighbors are searched randomly and the first neighbor better than the current solution is chosen. (With FIR we disable the tabu list for the initial solution because randomization makes the algorithm follow different paths at every execution, and we generate at most *max–n* neighbors.) If there is no neighbor better than the current solution or if the path length is longer than *cutoff* and there exists an unvisited initial solution, then a restart is conducted (Steps 4 and 5).

**Algorithm 2** Local search on best response vertices (LS–v)

1: randomly choose an unvisited initial solution $s$ from $\tilde{\mathbf{x}}_i = \mathbf{0}$ by pivoting
2: **repeat**
3:     choose neighbor $\theta' \in N(\theta)$ with $f(\theta') < f(\theta)$ and assign $\theta = \theta'$
4:     **if** there is no $\theta'$ or the path is longer than *cutoff* **then**
5:         go to Step 1
6: **until** $f(\theta) > 0$
7: return current solution

The advantages of rrL are that it can traverse solutions that are not in LH/L paths and that the number of neighbors is low (w.r.t. local search on supports). The drawback is that the algorithm is incomplete.

## 5 Anytime iterative perturbation over paths

Given a two–player game with utilities $(U_1, U_2)$ and a game with $(U'_1, U'_2)$ where every $U'_i$ is obtained by perturbing each payoff of $U_i$ with an arbitrary probability distribution over $[-\delta, \delta]$, it has been shown that an NE for $(U'_1, U'_2)$ is an $\epsilon$–NE for $(U_1, U_2)$ with $\epsilon \leq 4\delta$ [5]. We argue that the bound is tighter:

**Theorem 1.** *Given a two–player game with $(U_1, U_2)$ and a game with $(U'_1, U'_2)$ where every $U'_i$ is obtained by perturbing each payoff of $U_i$ with an arbitrary probability distribution over $[-\delta, \delta]$, an NE $(\hat{x}_1, \hat{x}_2)$ for*

$(U'_1, U'_2)$ is an $\epsilon$–NE for $(U_1, U_2)$ with $\epsilon \leq 2\delta$.

*Proof* Call $x_1^*$ the best response of agent 1 to $\hat{x}_2$ of agent 2 for the game $(U_1, U_2)$. By definition, $||U_i - U'_i||_\infty \leq \delta$. We can compute the upper bound over the expected utility loss of agent 1: $\epsilon = x_1^* U_1 \hat{x}_2 - \hat{x}_1 U_1 \hat{x}_2 \leq x_1^* U_1 \hat{x}_2 - x_1^* U'_1 \hat{x}_2 + \hat{x}_1 U'_1 \hat{x}_2 - \hat{x}_1 U_1 \hat{x}_2 \leq |x_1^* U_1 \hat{x}_2 - x_1^* U'_1 \hat{x}_2| + |\hat{x}_1 U'_1 \hat{x}_2 - \hat{x}_1 U_1 \hat{x}_2| \leq x_1^* ||U_1 - U'_1||_\infty \hat{x}_2 + \hat{x}_1 ||U'_1 - U_1||_\infty \hat{x}_2 \leq ||U_1 - U'_1||_\infty + ||U'_1 - U_1||_\infty \leq 2\delta$. The same reasoning can be applied to agent 2, obtaining the same upper bound. Hence, the theorem is proved. □

We exploit this result to produce a simple anytime algorithm (Algorithm 3) based on the idea that an hard game could become easy after perturbed. The algorithm iteratively applies LH starting from a perturbed game obtained with a large perturbation (i.e., $\delta = 0.125$) and then exponentially reducing the perturbation at each step where LH has found an NE.

**Algorithm 3** Anytime iterative perturbation over paths (ip-LH)

1: set $k = 3$
2: **while** *deadline* has not been reached **do**
3:  set $\delta = 1/2^k$
4:  apply $\delta$–uniform perturbation on $(U_1, U_2)$
5:  apply LH to perturbed $(U_1, U_2)$
6:  set $k = k + 1$
7: return current solution

## 6 Experimental evaluation

### 6.1 Experimental setting

We implemented LH, L, rrLH, rrL and ip–LH in C in two different versions: one with floating–point and one with arbitrary–precision integer arithmetic by means of the GMP library[2]. We implemented LS–v, PNS and LS–PNS in C calling CPLEX via AMPL to solve LPs, while MIP Nash directly in AMPL[3] with CPLEX[4]. All the algorithms that use pivoting techniques are optimized with the revised technique [15] and those using GMP with integer pivoting [23] to save compute time. (Revised technique allows us to store and perform pivoting on a reduced tableau, obtaining a relevant improvement in the performance of the algorithms.) We conducted the experiments on a UNIX computer with 2.33GHz CPU, 16GB RAM and kernel 2.6.24.

Using GAMUT [17], we generated 500 instances of CovariantGame–Rand, GraphicalGame–RG and PolymatrixGame–RG games (being the hardest classes w.r.t. PNS, LH and MIP Nash) and 10 instances of all the other classes with $m_1 = m_2 = m \in \{5, \ldots, 150\}$ and a step of 5. We also generated an instance of SGC games with $m \in \{3, \ldots, 99\}$ and a step of 4 as prescribed in [19]. In SGC games, all equilibria have a medium number of pure strategies in their supports. Finally, we generated one instance of HtSG (its generation not being random) with $m \in \{2, \ldots, 18\}$ and a step of 2 as prescribed in [20]. In HtSG, all the LH paths are exponentially long.[5]

### 6.2 Numerical instability with floating point precision

We evaluated numerical instability of LH and L with floating–point precision for different game classes, comparing their results to those obtained with arbitrary precision. With GAMUT game classes, L presents numerical instability (the algorithm goes in ray termination) even with small games ($m = 30$) for all the game classes, while LH presents instability with a limited number of game classes (not recognizing equilibria or finding non–NE solutions). With long double precision, LH performs a wrong path on 96.1% of 100x100 TravelersDilemma instances and on 22.75% of 100x100 WarOfAttrition instances. More available tricks could be used to diminish the numerical instability than the ones used in our implementation, but the result would be the same: LH and L are intrinsically unstable. Using other tricks we can increase the size of the games at which instability appears. However, the relative comparison of the game classes in terms of arbitrary precision stands. On HtSG, arbitrary precision arithmetic is required even during the game instance generation. We generated game instances using both the trigonometric and the non–trigonometric moment curves as described in the appendix of [20]. We counted, using the *lrs* algorithm [1], the number of vertices of the polyhedron associated with the game generated with floating point precision and compared it to that obtained with arbitrary precision. When at least one vertex is lost, the polyhedron loses its hardness characteristics and at least a short path appears. Tab. 1 shows that arbitrary precision is required for instances with $\geq 14$ actions per agent.

Table 1: Percentage of lost vertices in HtSG due to floating–point precision with different moment curves.

| Actions per agent | | | | | |
|---|---|---|---|---|---|
| 6 | 8 | 10 | 12 | 14 | 16 |
| non–trigonometric | | | | | |
| 0% | 0% | 0% | 0% | 1% | 29% |
| trigonometric | | | | | |
| 0% | 64% | 72% | 95% | 92% | 82% |

### 6.3 Path distribution

We ran LH along every possible path with all the 100x100 (99x99 with SGC) instances of our experimental setting (except HtSG), measuring the length of the paths and deriving their distribution for each game class. Almost all the distributions are *fat–tailed* [10]. These distributions present a lot of data points in the tail, showing that the performance of an algorithm may vary dramatically from run to run. Formally, a distribution is fat–tailed if its *kurtosis* ($\mu_4/\mu_2^2$, where $\mu_j$ is the $j$–th moment) is larger than 3 (i.e., the kurtosis of a standard normal). All the game classes have a kurtosis larger then 3, except for DispersionGame, MinimunEffortGame (the kurtosis is not defined since $\mu_2 = 0$ and $\mu_4 = 0$, all the paths having the same length), SGC (1.0), TravelersDilemma (1.8), and WarOfAttrition (2.39). As suggested in [10], with fat–tailed distributions, random restarts may drastically improve performance. Our experiments, discussed below, confirm this.

---
[2] http://gmplib.org/.
[3] http://www.ampl.com.
[4] http://www-01.ibm.com/software/integration/optimization/cplex-optimizer/.

[5] Although HtSG square instances can be easily solved by support enumeration algorithms, the equilibrium being unique and fully mixed, non–square games can be built such that support enumeration algorithms require exponential time.

Although almost all the classes present fat–tailed distributions, the performance of LH varies greatly across them. We ran LH with instances of different sizes and bucketed the game classes into five groups:

1. (DispersionGame, MinimumEffortGame) The length of all the paths is constant ($= 2$) with game size.
2. (BidirectionalLEG–CG/RG/SG, CovariantGame–Pos, LocationGame, UniformLEG–CG/RG/SG, WarOfAttrition) The average and maximum path length tends to a constant value ($< 10$) as game size increases.
3. (SGC, TravelersDilemma) The average path length increases linearly in game size.
4. (BertrandOligopoly, CovariantGame–Rand/Zero, GraphicalGame–RG/Road/SG/SW, PolymatrixGame–CG/RG/Road/SW, RandomGame) These games have an exponential growth of the average number of steps with the game size, but there can be some paths with polynomial length. Fig. 1 reports, for two classes, the average number of steps and the pertinent *box–plot* diagram: *median*, 1–st and 3–rd *quartiles* (dashed), *max* and *min* (dotted).
5. (HtSG). All the paths are exponentially long.

We produced the same analysis for L by randomly generating $m$ initial solutions. All the classes present the same behavior they have with LH — except DispersionGame, which, when solved with L, belongs to Group 2. HtSG preserves, with L, the same characteristic exhibited with LH: the length of the shortest paths grows exponentially.

### 6.4 Finding an NE by rrLH

Groups 1–3 are easy even without resorting to random restarts (these games are easy also with other algorithms, e.g. PNS). Thus, we just briefly summarize the main results omitting details. Instances of

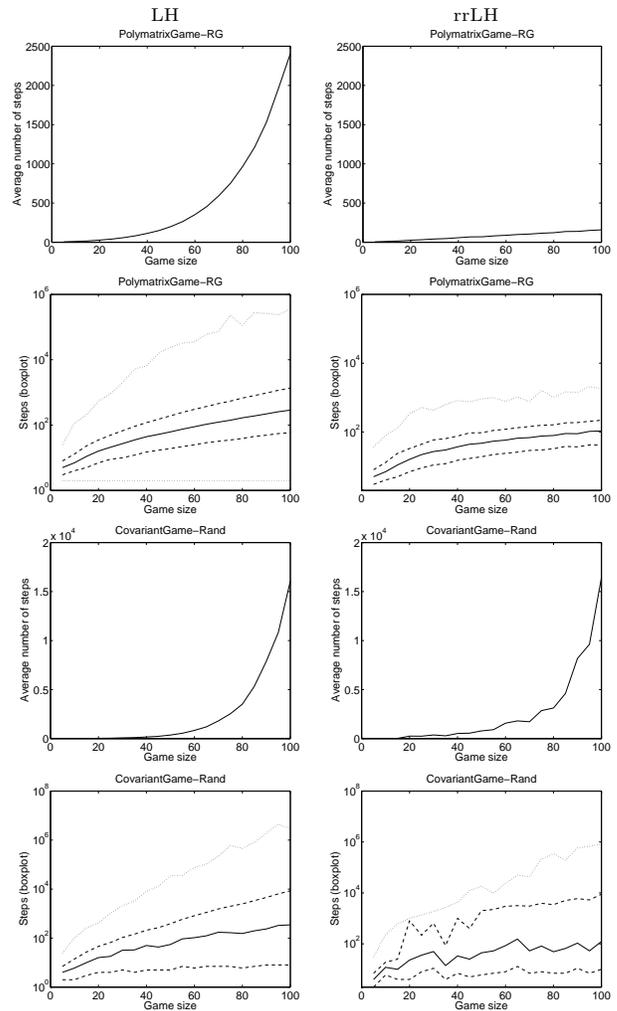

Figure 1: Comparison between LH (left) and rrLH (right) in terms of average number of steps and box–plot (median, 1–st/3–rd quartiles with dotted lines, min/max with dashed lines) of PolymatrixGame–RG and CovariantGame–Rand.

Groups 1 and 2 are solved by LH with a very small number of steps even in the worst cases (the compute times are less than one second). With Group 3, LH requires a linear compute time. With SCG it has already been experimentally demonstrated that LH outperforms MIP Nash and PNS [19]. With TravelersDilemma, LH (with arbitrary precision) is essentially the unique applicable algorithm — the commercial LP solvers do not support arbitrary precision, while a preliminary evaluation of GLPK with arbitrary precision led to compute times two orders of magnitude longer.

Random restarts play a crucial role with Group 4. These games are hard (in average) when solved with LH (and the other known algorithms: PNS, MIP Nash, LS–PNS). The application of rrLH with *cutoff* = 20 makes all the classes except for CovariantGame–Rand easy: all the executions of rrLH find an NE by the deadline and the average compute time is lin-ear the game size (the same for max and median), as shown in Fig. 1 for PolymatrixGame–RG. With CovariantGame–Rand, we observed that there is at least an instance whose shortest path grows in length exponentially (therefore, as shown in Section 3, no *cut-off*, even with iterative deepening, can lead to non–exponential compute times). As a result, the average number of steps is not smaller than that of LH.

Table 2: Termination percentage and compute times.

| rr–LH | | MIP Nash | | PNS | |
|---|---|---|---|---|---|
| BetrandOligopoly | | | | | |
| (100%) | 0.27 s | (100%) | 2.25 s | (100%) | **0.15 s** |
| CovariantGame–Rand | | | | | |
| **(80%)** | **176.61 s** | (10%) | 254.02 s | (20%) | 18.13 s |
| CovariantGame–Zero | | | | | |
| **(100%)** | **0.15 s** | (80%) | 155.79 s | (90%) | 3.75 s |
| GraphicalGame–RG | | | | | |
| **(100%)** | **0.16 s** | (10%) | 124.42 s | (90%) | 3.64 s |
| GraphicalGame–Road | | | | | |
| **(100%)** | **0.13 s** | (60%) | 98.32 s | (90%) | 4.47 s |
| GraphicalGame–SG | | | | | |
| **(100%)** | **0.15 s** | (40%) | 180.61 s | (60%) | 2.30 s |
| GraphicalGame–SW | | | | | |
| **(100%)** | **0.14 s** | (30%) | 155.28 s | (90%) | 3.42 s |
| PolymatrixGame–CG | | | | | |
| **(100%)** | **0.14 s** | (30%) | 213.77 s | (60%) | 3.27 s |
| PolymatrixGame–RG | | | | | |
| **(100%)** | **0.14 s** | (60%) | 160.80 s | (90%) | 1.14 s |
| PolymatrixGame–Road | | | | | |
| **(100%)** | **0.15 s** | (70%) | 208.48 s | (90%) | 2.51 s |
| PolymatrixGame–SW | | | | | |
| **(100%)** | **0.13 s** | (20%) | 73.73 s | (90%) | 2.43 s |
| RandomGame | | | | | |
| **(100%)** | **0.13 s** | (50%) | 160.14 s | (60%) | 4.80 s |

We compared rrLH with floating–point precision (arbitrary precision is averagely 56 times slower) to MIP Nash and PNS with 150x150 instances of Group 4 (LS–PNS is not useful for non–hard instances [3]). Tab. 2 shows the percentage of instances solved within 600 s and the compute time in seconds. rrLH dramatically outperforms the other two algorithms. More precisely, PNS terminates very quickly if there is an NE with very small support and takes exponential time otherwise. Call $\underline{s}$ the smallest support size of all the NEs of a game. PNS scans $O(n^{2\underline{s}})$ supports before finding an NE, instead we observed that rrLH makes $O(2n\underline{s})$ pivoting steps. Thus, as $n$ increases, PNS is inefficient even with small $\underline{s}$, instead rrLH scales well. MIP Nash performs radically worse than the others.

With HtSG, obviously, rrLH is not effective.

### 6.5 Finding an NE by rrL

The performance of rrL with blind random restarts are similar to rrLH, requiring for some classes (e.g. CovariantGame–Rand) a smaller number of steps but requiring more time per step. Given that all the other classes can be easily solved by applying rrLH except for CovariantGame–Rand and HtSG, we focused only on these two hard classes exploiting the main pecu-liarity of rrL: adopting a non–blind approach, trying to characterize good initial solutions and discarding those that are not promising.

We initially studied correlation between the values of $\epsilon$, $\epsilon_{WS}$, and $r$ of the initial solutions and the length of the paths with L, and we observed that no statistical correlation is present. Thus, such parameters cannot be used to identify short paths and discard those potentially long. Instead, we found statistical correlation for CovariantGame–Rand (and PolymatrixGame–RG) between the length of the L paths and the distance $(d_\infty)$ in $||\cdot||_\infty$ between the initial solution and the equilibrium (the steps reduce as $d_\infty$ increases), as shown in Fig. 2. Because the equilibrium is unknown, we can only use the vertices of the simplices as initial solutions (these are by definition the farthest points). However, with these initial solutions, L behaves like LH. This justifies why, although L allows potentially infinite paths, LH performs like L.

Given the difficulty of characterizing good initial solutions, we tried to characterize good paths by analyzing how some metrics ($\epsilon$, $\epsilon_{WS}$, $r$, $z_0$) vary during the paths. We designed a measure defined as $decr(z_0, h) = \frac{1}{z_{0_1}} \sum_{k=1}^{h} |z_{0_k} - z_{0_{k+1}}|$ that is equal to 1 when $z_0$ decreases monotonically and greater than 1 otherwise. Fig. 2 shows that with CovariantGame–Rand, $decr(z_0, h)$ is high on long paths. However, experimentally, this strategy resulted ineffective even fixing a small *cutoff*. Furthermore, there is no statistical correlation when $\epsilon$ or $\epsilon_{WS}$ or $r$ are used in place of $z_0$.

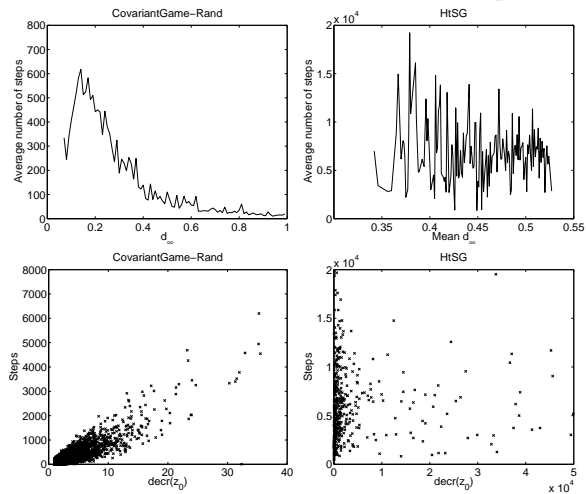

Figure 2: Relation between the measures $d_\infty$, $decr(z_0, m)$ and number of steps with 50x50 CovariantGame–Rand and 16x16 HtSG.

### 6.6 Finding an NE by LS–v

We mainly focused the evaluation of LS–v on CovariantGame–Rand, the other classes being easy and LS–v being not applicable to HtSG due to lack of arbitrary precision. We tuned the parameters

*cutoff* and *max–n* of FIR in LS–v as follows. We isolated 15 100x100 hard CovariantGame–Rand instances that cannot be solved by 600 s by PNS, MIP Nash or rrLH. We randomly chose 5 instances and tuned LS–v with $cutoff \in \{m, 2m, m^2, 2m^2\}$ and $max\text{–}n \in \{m/2, m, 2m, m^2/2\}$. The best configuration was $cutoff = m^2/2$ and $max\text{–}n = 2m^2$. With such configuration, the number of random restarts is very small. In practice, every solution has a better neighbor with high probability. Then we compared the performance of the three heuristics BI, FI and FIR (with its best configuration) with the other 10 hard instances (non–used for the tuning). The best heuristic was FIR with $\epsilon = 3.64 \cdot 10^{-4}$ by 600 s, while $\epsilon > 3 \cdot 10^{-3}$ with the other two. However, LS–v never found an NE with these 10 hard instances. In addition, we evaluated LS–v with generic (not necessarily hard) instances. Tab. 3 reports success probability and $\epsilon$ of the best solution found by 600 s. LS–v is outperformed by rrLH and rrL (that found an NE with GraphicalGame–RG and PolymatrixGame–RG with a probability of 100% and of 30% with 100x100 CovariantGame–Rand).

Table 3: Percentage of solved instances and average best $\epsilon$–NE found within 600 s with LS–v.

| Actions per agent | | |
|---|---|---|
| 60 | 80 | 100 |
| CovariantGame–Rand | | |
| (56%)9.34 · 10⁻⁵ | (63%)1.82 · 10⁻⁴ | (13%)1.80 · 10⁻⁴ |
| GraphicalGame–RG | | |
| (56%)1.32 · 10⁻⁴ | (38%)1.72 · 10⁻⁴ | (36%)7.83 · 10⁻⁴ |
| PolymatrixGame–RG | | |
| (60%)3.91 · 10⁻⁵ | (40%)3.95 · 10⁻⁴ | (30%)2.57 · 10⁻⁴ |

### 6.7 Approximating an NE

We compared LS–v to the anytime versions of the other algorithms, obtained by keeping track of the best $\epsilon$–NE during the pivoting, and ip–LH. Tab. 4 shows that LS–v is the best anytime algorithm (including PNS–anyT) for the 15 hard CovariantGame–Rand instances by an order of magnitude. It can be observed that ip–LH does not perform well with CovariantGame–Rand. This result shows that CovariantGame–Rand instances are stable, keeping to be hard even when perturbed.

We compared the anytime performance of LS–v and LH–anyT, these algorithms working on a similar solution space. Fig. 3 shows that LS–v found very good approximate solutions within a small number of steps and it outperforms LH–anyT for all setting of run time (the other heuristics provide similar results: LS–v finds very quickly a good approximate equilibrium).

We studied the $\epsilon$–NEs with 16x16 HtSG. These games were easy for ip–LH (with arbitrary precision), requiring less than 10 steps even with a perturbation of $10^{-10}$, returning $\epsilon \approx 10^{-12}$ (we cannot apply a perturbation smaller than $10^{-10}$ due to limits of GMP library). This shows that HtSG instances are highly unstable when perturbed. Therefore, there must exist another game class that is hard for LH and that is stable unless PPAD $\subseteq$ RP. Finally, LH–anyT provided best performance, returning $\epsilon \approx 10^{-34}$ within 600 s.

Table 4: Average best $\epsilon$–NE found within 600 s on 100x100 hard CovariantGame–Rand.

| LH–anyT | L–anyT | PNS–anyT | MILP–anyT |
|---|---|---|---|
| 6 · 10⁻³ | 2 · 10⁻² | 8 · 10⁻² | 3 · 10⁻³ |

| ip–LH | LS–v | LS–PNS |
|---|---|---|
| 1 · 10⁻² | **3 · 10⁻⁴** | 2 · 10⁻³ |

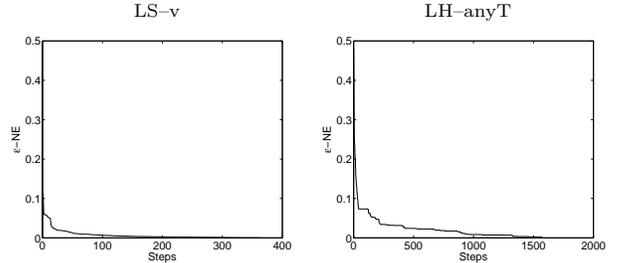

Figure 3: How the best $\epsilon$–NE found so far changes during two sample executions of LS–v and LH–anyT.

## 7 Conclusions and future research

The computation of an NE is a challenging task even with two agents. In this paper we present by and large the fastest algorithms for the problem to date. We also present other results about the problem. For many situations, arbitrary–precision arithmetic is necessary even with LH. Complementary pivoting with blind random restarts over paths is the best heuristic, linearizing the average compute time in game size and outperforming all our and state–of–the–art algorithms for all the benchmark game classes except CovariantGame–Rand and HtSG. We provide theoretical results that allow us to say that rrLH is asymptotically optimal among algorithms that randomize over LH paths. We did not find any metric to characterize good L paths and therefore to have non–blind random restarts. Local search guided by the minimization of $\epsilon$ exhibits worse performance in finding exact equilibria, but it is the best in approximating hard instances, except for HtSG. HtSG games are unstable in the sense that they can be easily approximated by introducing a very small perturbation; therefore there must exist an alternative hard instance for LH that is stable (CovariantGame–Rand is a possible candidate, it being stable). In the future, we will extend our algorithms to games with more than two agents and isolate hard stable instances for each algorithm.

### Acknowledgements

Tuomas Sandholm was funded under NSF grants IIS-0964579, CCF-1101668, and IIS-0905390.